# Time-Domain Analysis of Surface Acoustic Wave Magnetoelectric Sensors

Mohsen Samadi [a,*], Henrik Wolframm [b], Felix Weisheit [c], Dirk Meyners [c], Eckhard Quandt [c], Michael Höft [b], Martina Gerken [a,*]

[a] *Chair for Integrated Systems and Photonics, Department of Electrical and Information Engineering, Kiel University, 24143 Kiel, Germany*

[b] *Chair of Microwave Engineering, Department of Electrical and Information Engineering, Kiel University, 24143 Kiel, Germany*

[c] *Chair for Inorganic Functional Materials, Department of Materials Science, Kiel University, 24143 Kiel, Germany*

[*] *Corresponding author at: Chair for Integrated Systems and Photonics, Department of Electrical and Information Engineering, Kiel University, 24143 Kiel, Germany.*

*E-mail address: mosa@tf.uni-kiel.de (M. Samadi), mge@tf.uni-kiel.de (M. Gerken)*

**Abstract:** The transient response of surface acoustic wave sensors is important for understanding their dynamic behavior, propagation delay, and response to time-varying signals. In this work, the time-domain propagation characteristics of a magnetoelectric SAW sensor based on a multilayer Love-wave delay line are investigated using complementary numerical and experimental approaches. Time-domain analysis is employed to determine the delay time and to extract the phase and group velocities, while frequency-domain analysis provides an independent characterization of the wave dispersion. Both numerically and experimentally, the phase and group velocities are found to differ, reflecting the dispersion of the Love wave in the multilayer structure. At the operating frequency of 146 MHz, the simulated group velocity obtained from wave-field tracking agrees within 0.4% with the value determined from the frequency-domain dispersion relation. The simulated delay time and group velocity also show good agreement with experiment, with deviations of 2.7% in delay time and 0.9% in group velocity. These results demonstrate the consistency of the time- and frequency-domain analyses and provide a quantitative characterization of the transient propagation behavior of SAW sensors.

## 1. Introduction

Precise measurement of magnetic fields is crucial for a wide range of applications, including medical diagnostics and industrial monitoring [1–9]. Among various sensing technologies, surface acoustic wave (SAW) sensors are widely used because of their high sensitivity and capability to detect minute magnetic field variations over a broad frequency range [10–16]. These sensors typically rely on the $\Delta E$ effect, whereby the mechanical stiffness of a magnetostrictive layer changes in response to an external magnetic field [17–20]. This variation in material properties modulates the velocity of the propagating SAW, which can be detected as a phase shift at the output. The influence of magnetic-field-dependent material properties on propagation characteristics has been investigated in SAW devices [21–24].

SAW sensors commonly utilize a piezoelectric substrate such as ST-cut quartz or a piezoelectric thin film, such as aluminium scandium nitride (AlScN) thin films, to generate and detect acoustic waves via interdigital transducers (IDTs) [11,15]. A guiding layer is often deposited on the piezoelectric material to confine the acoustic energy near the surface, thereby enhancing the interaction between the SAW and the magnetostrictive layer. Recent studies have explored various approaches to enhance the sensitivity of SAW magnetic field sensors, such as exploiting the slow-wave phenomenon in phononic crystals (PnCs) [25,26] or strong magnon–phonon coupling in acoustic cavities [27].

Several analytical and numerical approaches have been employed to describe the propagation characteristics of SAW delay-line devices [28]. In particular, the finite element method (FEM) has been widely used to model Love-wave delay-line devices [14,29–31]. Time-domain analysis provides information complementary to conventional frequency-domain characterization of SAW sensors. While frequency-domain analysis is well suited to determining resonance and dispersion characteristics, time-domain simulation provides access to the spatiotemporal evolution of the acoustic field, allowing wave propagation to be examined through the motion of wave peaks and wavefronts. This is particularly relevant for multilayer Love-wave structures, where the spatial distribution of acoustic energy influences the relationship between phase and group velocities. The temporal evolution of the output signal also provides a measure of the propagation delay through the delay line, which is related to the group velocity. The delay time further determines the temporal averaging inherent to a delay-line sensor and consequently influences its response to time-varying input signals [32]. Thus, time-domain analysis provides a framework for relating the propagation characteristics of the SAW to the dynamic response of the sensor.

In this work, we present a systematic time-domain analysis of a magnetoelectric SAW sensor based on a multilayer Love-wave delay-line configuration. The present study focuses on the propagation characteristics and transient response of the sensor rather than its magnetic sensing performance. A 3D FEM model is developed to simulate the transient response to sinusoidal electrical excitation and to directly extract the delay time, phase velocity, and group velocity from the simulated time-dependent wave field. While FEM models have previously combined frequency- and time-domain analyses [31] to determine the phase velocity and coupling coefficient of non-dispersive SAWs, we extract both velocities directly from the transient wave field of a layered, dispersive Love-wave delay line. The results are compared with those obtained from frequency-domain simulations and further validated experimentally.

## 2. Theoretical Model

### 2.1. Equations

The electromechanical response of the SAW device is described by solving a system of coupled differential equations. The mechanical behavior is governed by the equation of motion, which links the displacement vector **u** to the divergence of the mechanical stress tensor **σ** [33,34]:

$$\nabla . \boldsymbol{\sigma} = \rho \frac{\partial^2 \mathbf{u}}{\partial t^2}, \tag{1}$$

where $\rho$ denotes the material density. Assuming that free charges and eddy currents can be neglected, the electrostatic behavior follows Maxwell's equation in its electrostatic form [35]:

$$\nabla . \mathbf{D} = 0, \tag{2}$$

where **D** is the electric flux density vector. The electric field **E** is obtained from the spatial gradient of the electric potential *V*, expressed as:

$$\mathbf{E} = -\nabla V. \tag{3}$$

The mechanical and electrostatic responses are coupled through the constitutive piezoelectric equations in the stress-charge form [36]:

$$\boldsymbol{\sigma} = \mathbf{C}\boldsymbol{\varepsilon} - \mathbf{e}^{\mathrm{T}}\mathbf{E}, \tag{4}$$

$$\mathbf{D} = \mathbf{e}\boldsymbol{\varepsilon} + \boldsymbol{\varepsilon}_{\mathbf{el}}\mathbf{E}. \tag{5}$$

In this formulation, $\boldsymbol{\varepsilon}$, **C**, **e**, and $\boldsymbol{\varepsilon}_{\mathbf{el}}$ correspond to the strain tensor, the elastic stiffness tensor, the piezoelectric coupling tensor, and the dielectric permittivity tensor, respectively. The material parameters are provided in Appendix B.

### 2.2. Simulation Approach

To simulate the frequency- and time-domain responses of the SAW sensor, an electromechanical model was developed in COMSOL Multiphysics® [37], as shown in Fig. 1. The model is built on an ST-cut quartz substrate, which serves as the piezoelectric medium for the generation and propagation of SAWs. The excitation region consists of IDT electrodes in a double-finger configuration with a pitch of $p$=28 μm patterned on the surface of the quartz substrate. As illustrated in the left inset of Fig. 1, the electrodes are made of Au and embedded in a $SiO_2$ guiding layer. When an alternating electrical signal is applied to the IDT electrodes, electrical energy is converted into mechanical deformation through the inverse piezoelectric effect, generating a SAW that propagates along the $x$-axis on the surface of the delay line. As shown in the right inset, the delay line comprises the ST-cut quartz substrate coated with a $SiO_2$ guiding layer and a FeCoSiB magnetostrictive layer. The fabrication procedure and device dimensions are described in detail in Appendix A.

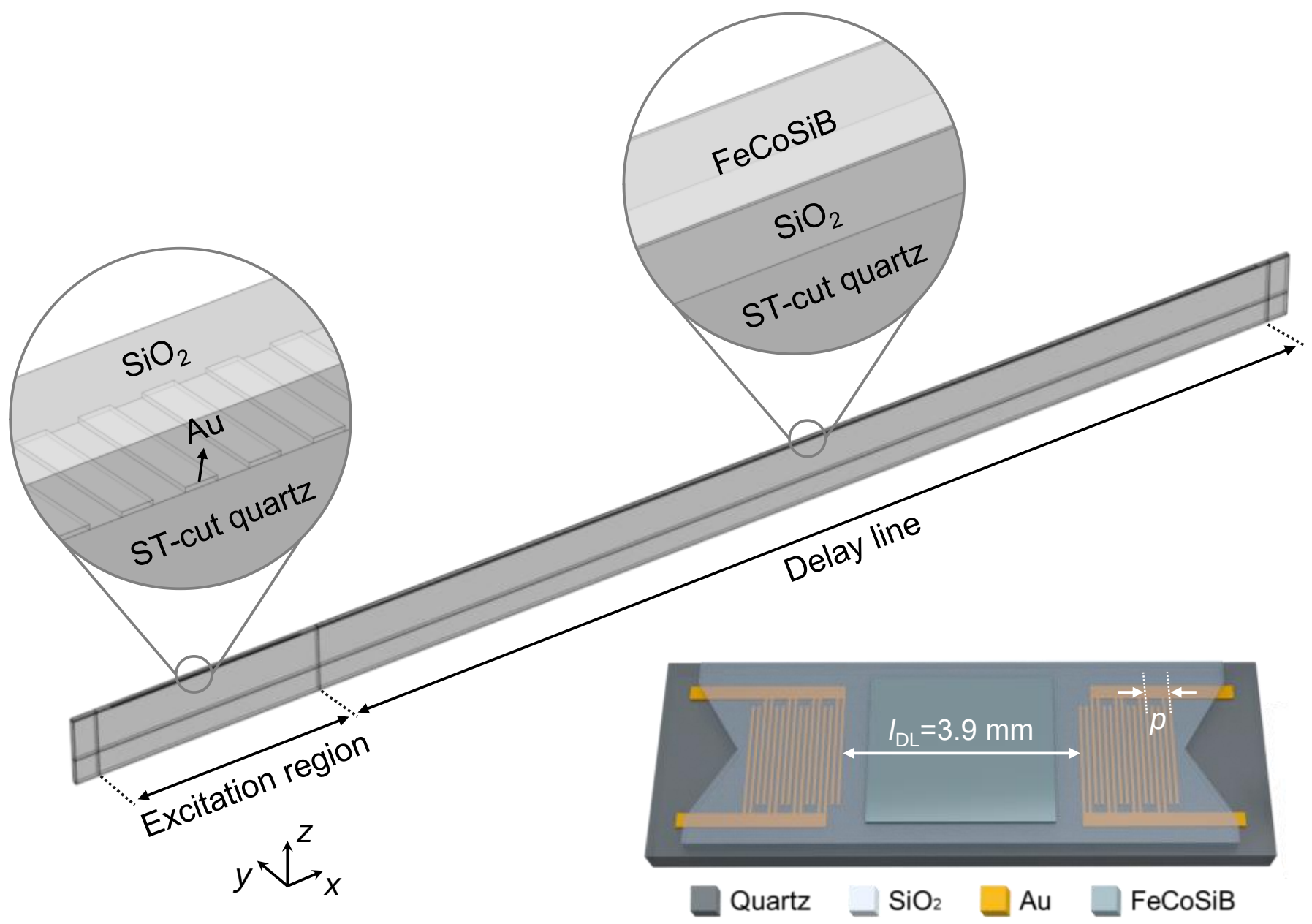

**Fig. 1.** Schematic of the SAW magnetoelectric sensor and the corresponding FEM model. The excitation region consists of Au IDT electrodes patterned on an ST-cut quartz substrate and embedded in a $SiO_2$ guiding layer (left inset). The generated SAW propagates along the $x$-axis on the surface of the delay line, which comprises the ST-cut quartz substrate, the $SiO_2$ guiding layer, and a magnetostrictive layer of FeCoSiB (right inset). Periodic boundary conditions are applied along the $y$-axis to emulate infinitely long IDT fingers, while perfectly matched layers (PMLs) are implemented in the $x$ and $z$ directions to suppress reflections at the model boundaries. The delay line length, $l_{DL}$=3.9 mm, is defined as the edge-to-edge distance between the input and output IDTs.

The simulation domain is limited to half of the IDT pitch ($p$/2=14 μm) along the $y$-axis to reduce the model size and computational cost. Periodic boundary conditions are applied in the $\pm y$ directions to emulate infinitely long IDT fingers. Perfectly matched layers (PMLs) are implemented in the $x$ and directions to suppress unwanted reflections from the domain boundaries. Furthermore, the substrate thickness is chosen to be sufficiently large to ensure that the displacement field becomes negligible at the bottom boundary. The output IDT is omitted from the model, as its inclusion would increase the computational complexity without providing additional insight into the wave propagation investigated in this study.

The simulations were performed in two steps. First, a frequency-domain study was conducted to determine the resonance frequency of the SAW at steady state. This was achieved by applying alternating electrical signals with frequencies ranging from 100 to 200 MHz to the IDT and calculating the average displacement across the surface of the delay line. The resonance frequency of the SAW corresponds to the frequency at which the average displacement reaches its maximum. Second, a time-domain study was carried out to investigate the temporal response of the sensor. For this purpose, a sinusoidal signal with a given frequency $f$ was applied to the IDT. The resulting output displacement was then calculated at time steps of $T/8$, where $T=1/f$ is the excitation period, in the range between $t_i=0$ and $t_f=2$ μs. The selected simulation duration is sufficiently long to ensure that the sensor reaches a steady state and that the transient response is fully captured.

## 3. Simulation Results

### 3.1. Frequency-domain Simulations

To obtain the SAW spectrum and determine its center frequency, frequency-domain simulations were performed over the range of 100–200 MHz. The displacement amplitude, $(u_x^2+u_y^2+u_z^2)^{1/2}$, corresponding to the magnitude of the three-dimensional displacement vector $\vec{u} = (u_x, u_y, u_z)$, was calculated, averaged over the surface of the delay line, and normalized to its maximum value. The simulation results, shown in Fig. 2a, indicate a SAW with a maximum amplitude at $f$=146 MHz, which matches the response of sensors with the same design [38] and is further confirmed by the experimental results shown in Fig. 6a. Decomposition of the displacement vector into its components reveals that shear-horizontal oscillations along the $y$-axis are dominant, whereas $u_x$ and $u_z$ are more than 60 dB lower than $u_y$. Consequently, their contribution to the average displacement is negligible.

When the sensor is operated at the resonance frequency ($f_{ex}$=146 MHz), a pure Love wave is generated that propagates along the $x$-axis, with shear-horizontal oscillations along the $y$-axis confined to the surface of the delay line (Fig. 2b). At resonance, the excitation period $T_{ex}=1/f_{ex}$ is well matched to the temporal periodicity of the IDT, $T_{IDT}=p/v_p$, where $p$=28 μm is the spatial periodicity of the IDT and $v_p$ is the SAW phase velocity. This ensures coherent excitation by the individual IDT electrodes, resulting in a higher displacement amplitude along the delay line. In contrast, excitation away from the SAW resonance leads to a mismatch between the excitation period and the temporal periodicity of the IDT. Consequently, the contributions from the individual IDT electrodes become less coherent, resulting in a lower average displacement amplitude. As illustrated in Fig. 2c, when the SAW is excited at $f$=143 MHz, the increased phase mismatch significantly reduces the excitation efficiency and consequently limits the buildup of the SAW along the delay line.

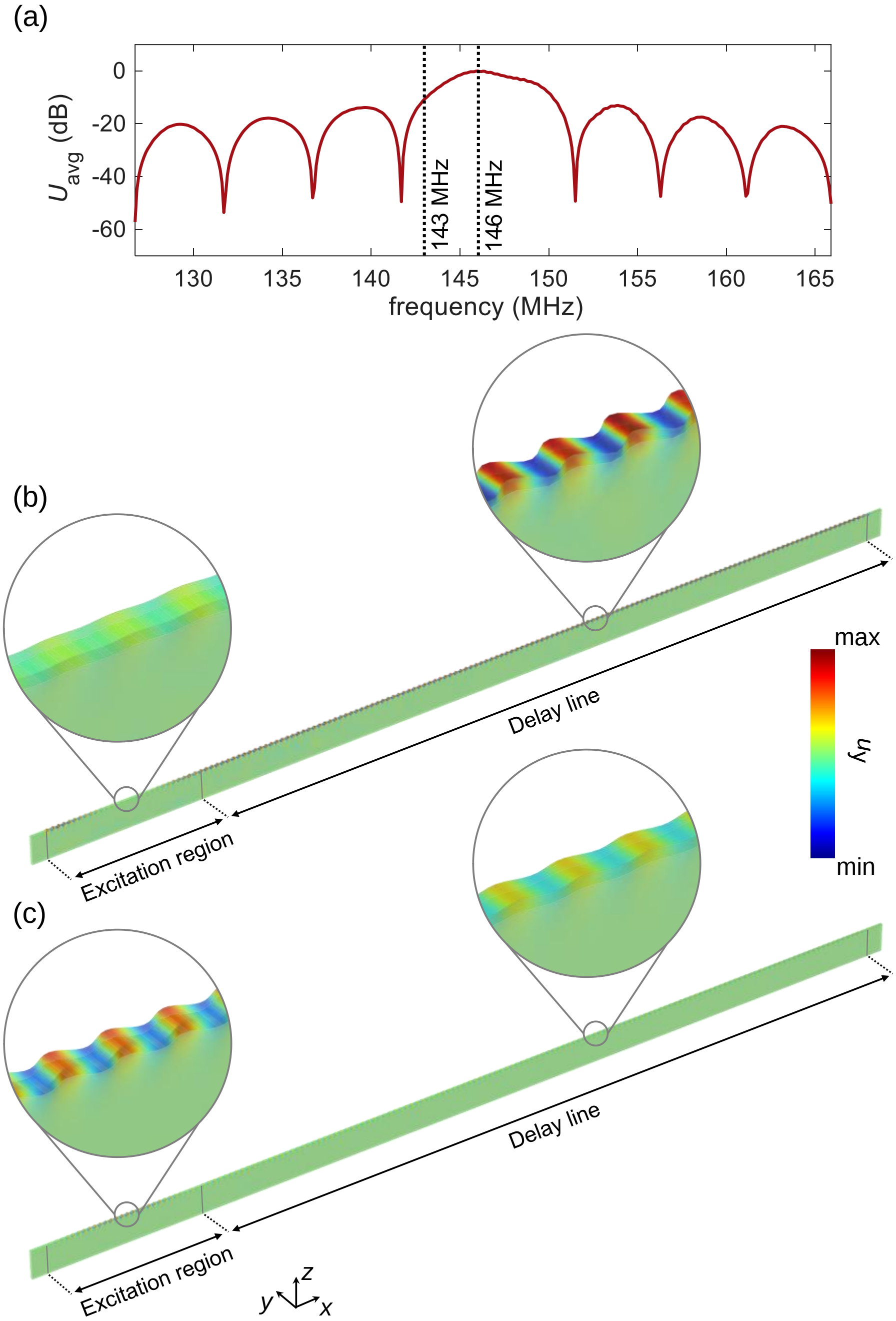

**Fig. 2.** (a) Displacement amplitude, $(u_x^2+u_y^2+u_z^2)^{1/2}$, averaged over the surface of the delay line and normalized to its maximum value at $f$=146 MHz. (b, c) Simulated shear-horizontal displacement field ($u_y$) distribution along the SAW delay line at different excitation frequencies: (b) At the resonance frequency ($f_{ex}$=146 MHz), coherent excitation produces a well-defined Love wave with strong shear-horizontal displacement along the $y$-axis and efficient propagation along the $x$-axis. (c) At an off-resonance frequency ($f_{ex}$=143 MHz), phase mismatch reduces the excitation efficiency, resulting in weaker displacement and limited SAW buildup along the delay line. Insets show magnified views of the displacement field in the excitation and delay-line regions.

Previous studies have shown that Love waves propagating in layered structures exhibit dispersive behavior, where their phase velocity depends on their wavelength [39,40]. We investigated the dispersion characteristics of the Love wave in the present sensor by numerically calculating its center frequency at various wavelengths. The wavelength was varied over the range $\lambda$=9.1-91 μm, corresponding to a normalized wavelength range of $\lambda/h_{GL}$=2-20, where $h_{GL}$=4.55 μm is the guiding layer thickness. The phase velocity was defined as $v_p=\lambda f_0$, where the wavelength $\lambda$ is determined by the IDT spatial periodicity $p$ and $f_0$ is the center frequency. The resulting dispersion curve, shown in Fig. 3, reveals an inverse relationship between phase velocity ($v_p$ , blue circles) and wavelength. At longer wavelengths, the Love wave penetrates deeper into the quartz substrate, which possesses a higher shear-wave velocity, resulting in higher phase velocities. As the wavenumber $k$ increases and the wavelength decreases, the acoustic energy becomes more confined within the $SiO_2$ guiding layer, whose shear-wave velocity is lower than that of the quartz substrate. Consequently, the Love-wave phase velocity decreases.

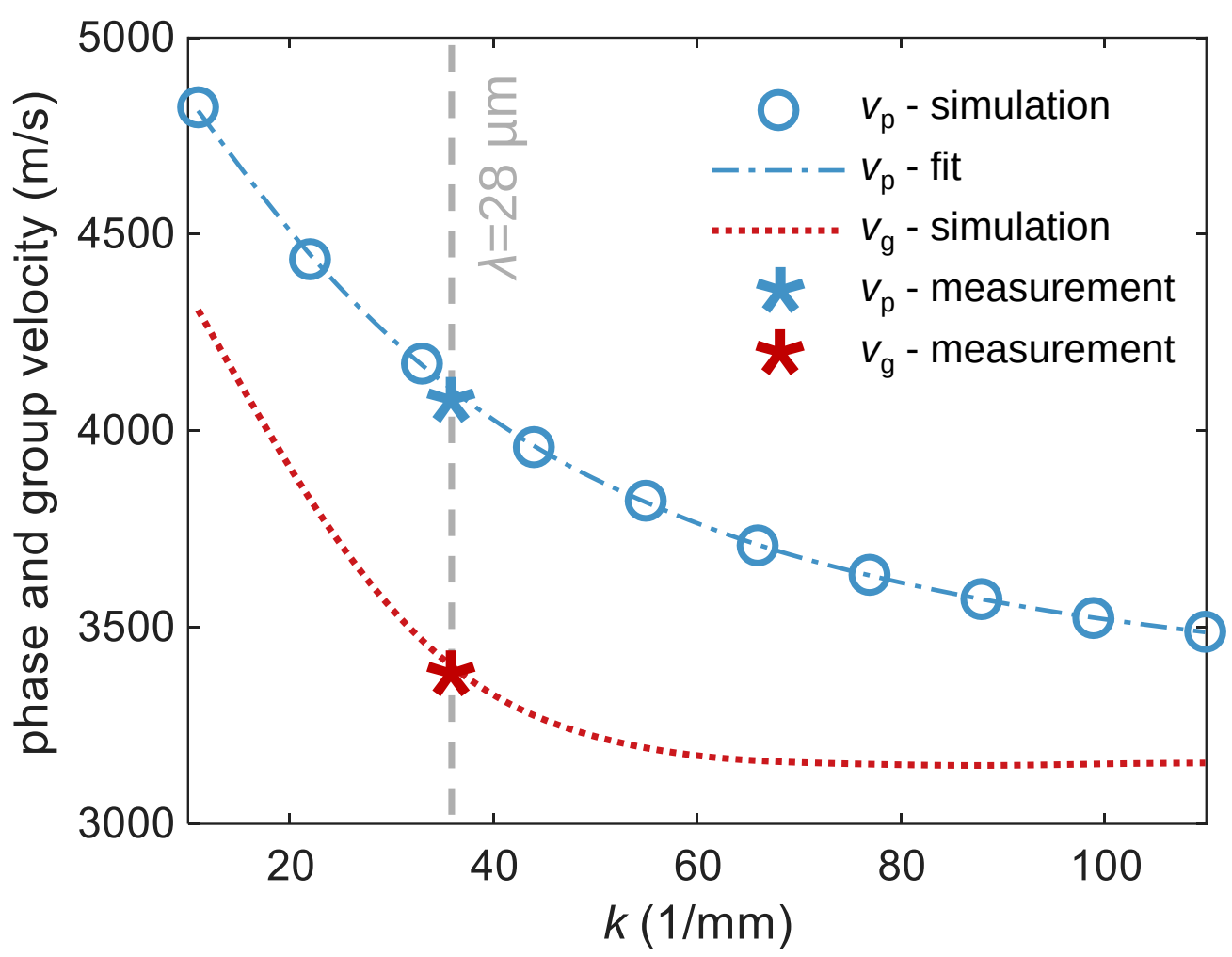

**Fig. 3.** The phase velocity ($v_p$ , blue circles) and the corresponding group velocity ($v_g$, red dotted line) as a function of the wavenumber $k=1/\lambda$. The wavelength is varied over the range $\lambda$=9.1-91 μm, corresponding to a normalized wavelength range of $\lambda/h_{GL}$=2-20, where $h_{GL}$=4.55 μm is the guiding layer thickness. The measured phase and group velocities at $\lambda$=28 μm are indicated by blue and red asterisks, respectively.

The raw phase velocity data were fitted using a cubic smoothing spline with a smoothing parameter of $p_s=1-1\times10^{-5}$, and the resulting fitted curve (blue dash-dotted line) was used to calculate the group velocity according to [39]

$$v_{\mathrm{g}} = v_{\mathrm{p}}\left(1 + \frac{h_{\mathrm{GL}}/\lambda}{v_{\mathrm{p}}}\frac{dv_{\mathrm{p}}}{d(h_{\mathrm{GL}}/\lambda)}\right), \tag{6}$$

which is shown as a function of the wavenumber $k$ in Fig. 3 (red dotted line). A substantial difference between the phase and group velocities is observed over the entire wavelength range. As discussed in detail in Sections 3.2 and 4, the phase and group velocities were estimated through time-domain simulations and experimental measurements at a wavelength of $\lambda$=28 µm. At this wavelength, which is indicated by a vertical dashed line in Fig. 3, the frequency-domain simulations yielded a phase velocity of $v_p \simeq 4105$ m/s and a group velocity of $v_g \simeq 3405$ m/s.

### 3.2. Time-domain Simulations

To investigate the transient response of the sensor, the frequency-domain simulations were followed by systematic time-domain simulations. A sinusoidal electrical signal, $V_{in}=A_0\sin(2\pi f_{ex}t)$, with an amplitude $A_0$=100 mV and an excitation frequency $f_{ex}$, was applied to the IDT structure. A smooth ramp function with a duration of $5T_{ex}$ was introduced to the sinusoidal electrical signal to avoid the broadband spectral components associated with an abrupt turn-on. The shear-horizontal displacement $u_y$ was probed as a function of time at the endpoint of the delay line, located 50 µm beyond the edge of the FeCoSiB layer. The delay line, defined as the edge-to-edge distance between the input and output IDTs, has a length of $l_{DL}$ =3.9 mm, as illustrated in Fig. 1. The displacement was sampled at time steps of $T_{ex}/8=1/(8f_{ex})$ over the interval from $t_i$=0, corresponding to the instant when the excitation signal is applied, to $t_f$=2 µs. To model the response detected by the actual output IDT, the point-probed displacement signal was convolved with a spatial sampling function representing the 25 finger pairs of the output transducer. This convolution accounts for the finite spatial extent of the receiving electrodes and their coherent summation of the propagating SAW, providing a more realistic representation of the output signal than single-point probing of the displacement.

The influence of the excitation frequency on the sensor's delay time was examined by applying electrical signals with slightly different frequencies to the input IDT. The resulting output displacement $U_{out}$ is presented in Fig. 4a for resonant excitation ($f_{ex}$=146 MHz, blue) and off-resonant excitation ($f_{ex}$=143 MHz, red). When the sensor operates at the resonance frequency ($f_{ex}$=146 MHz), the excitation is significantly more efficient, resulting in a larger output displacement for a given input signal amplitude. In contrast, the output signal decreases when the sensor operates away from the resonance frequency.

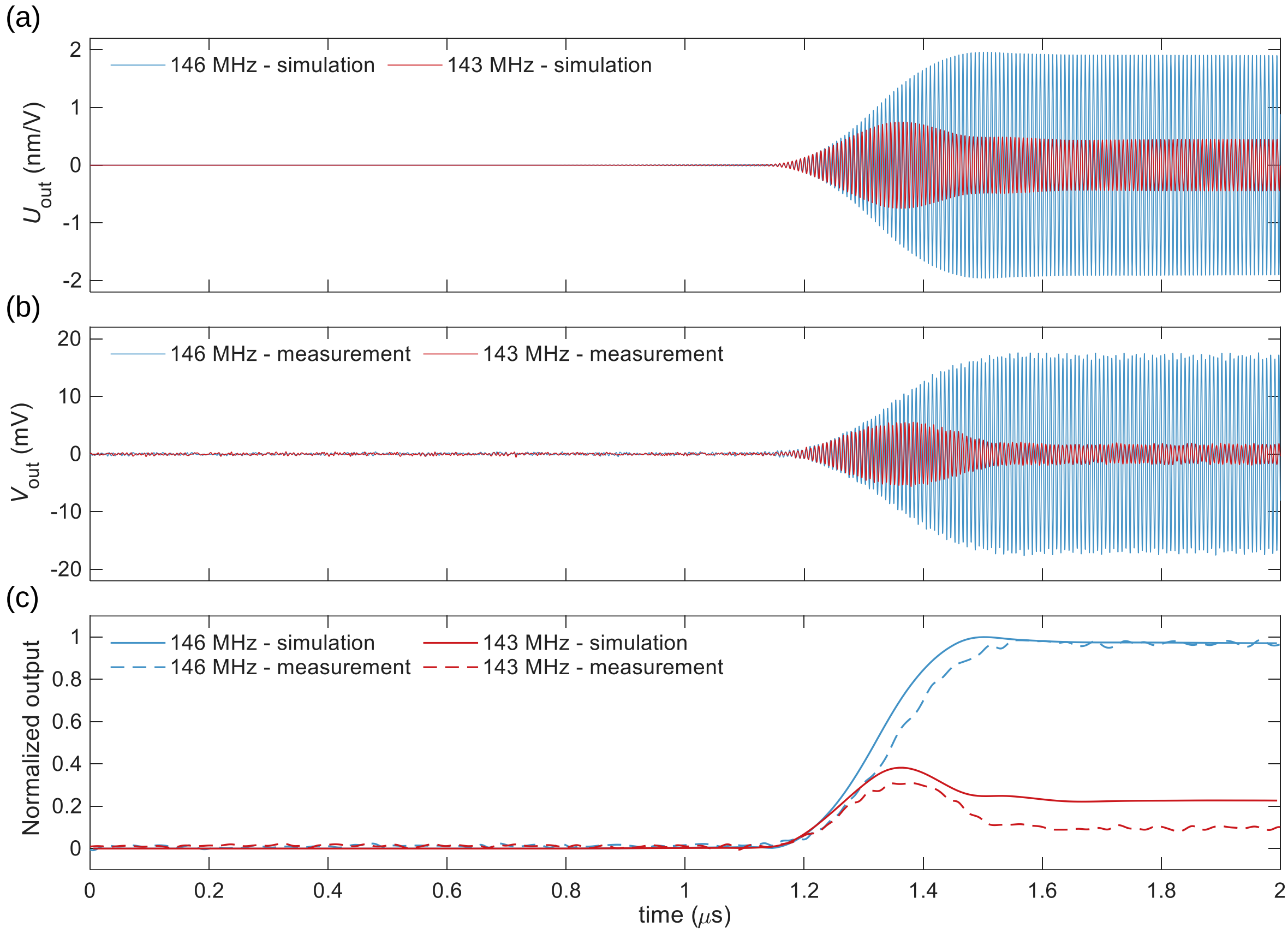

**Fig. 4.** (a) Simulated output displacement $U_{out}$ and (b) measured output voltage $V_{out}$ under resonant excitation ($f_{ex}$=146 MHz, blue) and off-resonant excitation ($f_{ex}$=143 MHz, red). The displacement was sampled at time steps of $T_{ex}/8=1/(8f_{ex})$ over the interval $t_i$=0 to $t_f$=2 μs and was normalized to the excitation signal amplitude. The measurement was performed at 500 MS/s, followed by interpolation to 5 GS/s. (c) Normalized peak values of the simulated (solid lines) and measured (dashed lines) output for $f_{ex}$=146 MHz (blue) and 143 MHz (red).

The delay time of the SAW sensor is defined as the time interval between the excitation instant ($t_i$=0) and the detection of the propagating SAW by the output IDT. This parameter characterizes the temporal response of the sensor and can be used to estimate the SAW group velocity. To determine the delay time from the time-domain simulations, the output signal is divided into three sections. The first section corresponds to the time interval during which no signal is detected and the amplitude is almost zero; the second corresponds to the transient state, when the amplitude changes from the initial state to a final steady state; and the third corresponds to the steady state, in which the amplitude remains nearly constant. The midpoint in time of the second (transient) section is defined as the delay time, $t_d$. At excitation frequencies of $f_{ex}$=146 MHz and 143 MHz, delay times of $t_d$≃1.321 μs and 1.314 μs were calculated, corresponding to group velocities of $v_g$≃3482 m/s and 3501 m/s. The group velocity was estimated by dividing the center-to-center distance between the input and output IDTs, $l_{cc}$=4.6 mm, by the corresponding delay time.

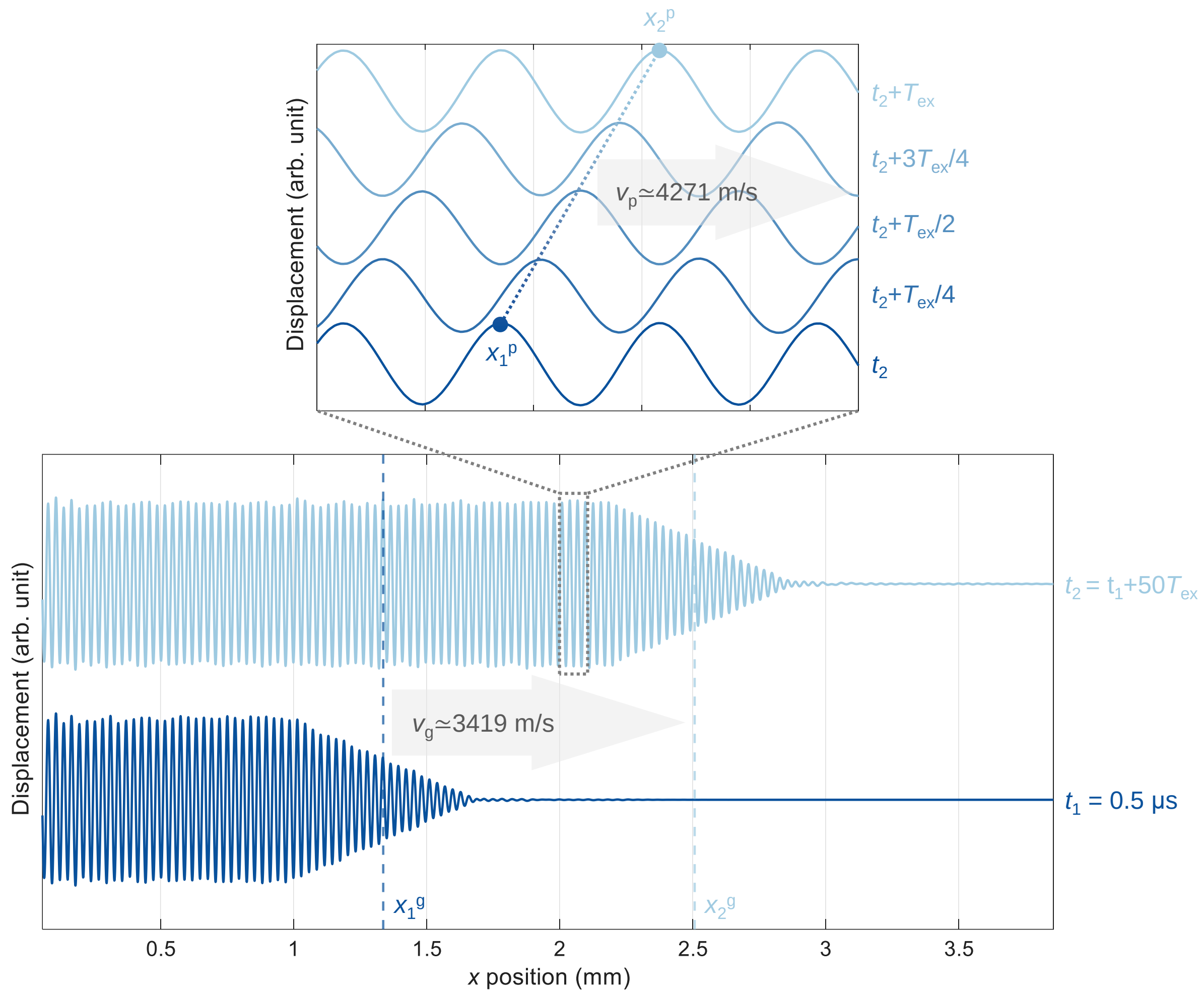

**Fig. 5.** Shear-horizontal displacement $u_y$ along the delay line at the excitation frequency $f_{ex}$=146 MHz, plotted at two instants $t_1$=0.5 μs (dark blue) and $t_2=t_1+50T_{ex}$ (light blue), where $T_{ex}=1/f_{ex}$ is the excitation period. The vertical dashed lines mark the $x$ position, at each time, where the wave holds 50% of its maximum amplitude. Group velocity of the wave is defined as the distance traveled by the wavefront ($x_2^g$-$x_1^g$) over the time duration $50T_{ex}$. The inset displays a zoomed-in view of the shear-horizontal displacement $u_y$ between $t_2$ (dark blue) and $t_2+T_{ex}$ (light blue). The dotted line tracks a single peak of the wave over the period $T_{ex}$. Phase velocity of the wave is defined as the average distance traveled by the wave peaks over the period $T_{ex}$.

In Fig. 5, shear-horizontal displacement ($u_y$) profiles along the delay line at the excitation frequency $f_{ex}$=146 MHz are plotted at two instants, $t_1$=0.5 μs (dark blue) and $t_2=t_1+50T_{ex}$ (light blue), where $T_{ex}$ is the excitation period. The vertical dashed lines indicate the positions $x_1^g \simeq 1337$ μm and $x_2^g \simeq 2508$ μm, where the displacement amplitude reaches 50% of its maximum value at times $t_1$ and $t_2$, respectively. The displacement of this amplitude threshold over the time interval $50T_{ex}$ is used to determine the group velocity, $v_g=(x_2^g-x_1^g)/50T_{ex} \simeq 3419$ m/s. The inset presents a magnified view of the displacement field between the time $t_2$ (dark blue) and $t_2+T_{ex}$ (light blue). The dotted line traces the movement of a wave peak from position $x_1^p$ to $x_2^p$ during the time interval $T_{ex}$. The phase velocity is estimated by calculating the distance traveled by each individual peak over one period, i.e., $(x_2^p - x_1^p)/T_{ex}$, and then averaging over all peaks along the delay line, yielding $v_p \simeq 4271$ m/s. Both the group and phase velocities are determined within the delay line region covered by the FeCoSiB layer.

The phase and group velocity values obtained from the time-domain simulations are compared with those derived from the frequency-domain analysis presented in Section 3.1. The group velocity differs by only 14 m/s (0.4%), whereas the phase velocity obtained from the time-domain simulations is 166 m/s (4%) higher than that obtained from the frequency-domain analysis. The larger difference in phase velocity values can be attributed to the assumption in the frequency-domain analysis that the SAW wavelength is equal to the IDT pitch. This assumption is only valid within the IDT region, whereas the wavelength differs in the delay line, where the phase velocity is calculated by tracking the wave peaks in the time domain, due to the different structural configuration and material composition. A comparable effect was experimentally observed by Müller et al. [23] in a similar device using time-resolved magneto-optical Kerr effect (MOKE) microscopy to directly image the propagating SAW and determine its wavelength and phase velocity, revealing a difference between the SAW wavelength in the FeCoSiB-covered delay line and that defined by the IDT periodicity.

As an additional consistency check, the group velocity calculated from the delay time is compared with that obtained from wavefront tracking. Under resonant excitation, the delay-time approach yields $v_g \simeq 3482$ m/s, while wavefront tracking gives $v_g \simeq 3419$ m/s, indicating a deviation of approximately 1.8%. The wavefront-tracking value is adopted as the reference for comparison with the experimental value, as it is not affected by the finite rise time of the transient.

## 4. Measurement Results

The time-domain response of the SAW sensor was experimentally measured at the operating frequencies of 146 MHz and 143 MHz, as shown in Fig. 4b. Details of the measurement method are provided in Appendix C. For direct comparison between the simulation and measurement results, the peak values of the simulated and measured output signals were extracted and plotted in Fig. 4c. The measured response shows good agreement with the simulation results, thereby providing experimental validation of the model.

Fig. 6 presents the measured output amplitude as a function of excitation frequency, together with the delay time determined from the time-domain response. Measurements were carried out at 10 kHz intervals over the frequency range from 142 to 151 MHz. The measured output amplitude in Fig. 6a was calculated as the ratio of the steady-state output amplitude to the excitation amplitude. The electrical bandwidth is defined as the frequency range in which the output amplitude remains within 6 dB of its maximum, i.e. above half the maximum amplitude. For the sensor investigated here, this yields an electrical bandwidth of 4.28 MHz. Within this bandwidth, the sensor exhibits resonant behavior, resulting in a monotonically rising signal during the settling process. Beyond this range, as shown in Fig. 4, an overshoot is observed due to the increased mismatch between the excitation period and the temporal periodicity of the IDT.

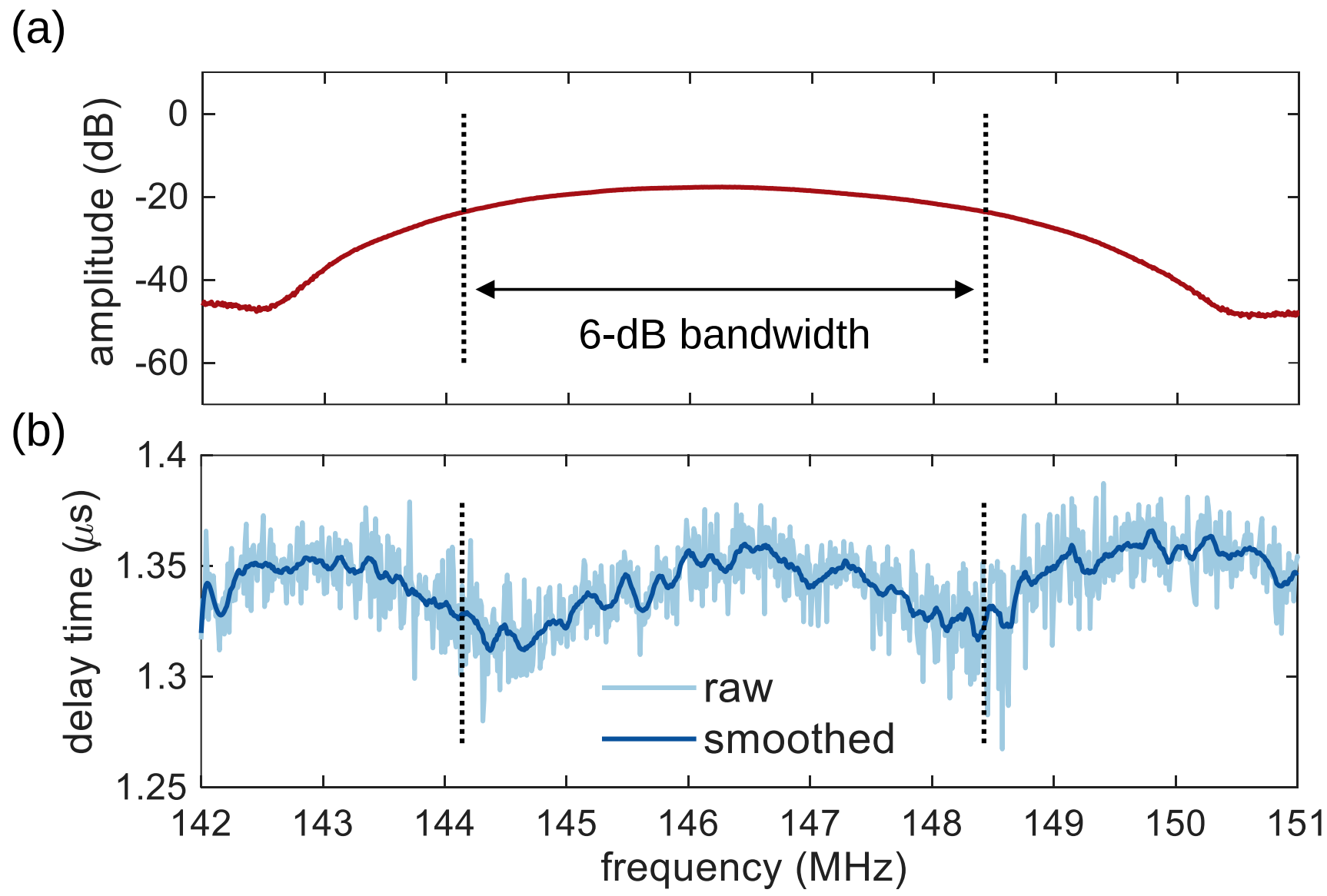

**Fig. 6.** (a) Measured output amplitude as a function of the excitation frequency. The sensor exhibits resonant behavior within the 6-dB bandwidth, i.e., between 144.14 MHz and 148.42 MHz. (b) Delay time as a function of the excitation frequency, measured in the time domain.

The same approach as that described in Section 3.2 was used to determine the delay time and group velocity from the time-domain measurements. At excitation frequencies of $f_{ex}$=146 MHz and 143 MHz, delay values of $t_d \simeq 1.357$ μs and 1.350 μs were measured, respectively. Under resonant excitation, a deviation of 36 ns (2.7%) from the time-domain simulations is observed. A group velocity of $v_g \simeq 3390$ m/s was obtained under resonant excitation, deviating by approximately 29 m/s (0.9%) from the time-domain analysis. The phase velocity was calculated using the same method as in the frequency-domain simulations, i.e., $v_p = \lambda f_0$, where $\lambda$=28 μm and $f_0$=146.3 MHz is the center frequency obtained from Fig. 6a, resulting in a measured phase velocity of $v_p \simeq 4096$ m/s. This value is approximately 9 m/s (0.2%) lower than the phase velocity estimated at the same wavelength using the frequency-domain simulation. The measured phase and group velocities are indicated by blue and red asterisks, respectively, in Fig. 3 for comparison with the frequency-domain results. A detailed comparison of the experimental, frequency-domain, and time-domain results is provided in Table 1.

**Table 1.** Comparison of simulation and measurement results

| Quantity (unit) | Measurement | Simulation | |
|---|---|---|---|
| | | Freq. domain | Time domain |
| Center frequency $f_0$ (MHz) | 146.3 | 146.6 ($\delta$=0.2%) | - |
| Phase velocity $v_p$ (m/s) | 4096 [a] | 4105 [a] ($\delta$=0.2%) | 4271 [b] ($\delta$=4.3%) |
| Group velocity $v_g$ (m/s) | 3390 | 3405 [c] ($\delta$=0.4%) | 3419 [d] ($\delta$=0.9%) |
| Delay time $t_d$ (µs) | 1.357 | - | 1.321 ($\delta$=2.7%) |

[a] $v_p=\lambda\cdot f_0$, where $\lambda=p$=28 µm.

[b] $v_p=\Delta x^p/T_{ex}$, where $\Delta x^p$ is the distance travelled by the wave peaks during one excitation period.

[c] Calculated from the dispersion relation (6), at $\lambda$=28 µm.

[d] $v_g=\Delta x^g/50T_{ex}$, where $\Delta x^g$ is the distance travelled by the wavefront over 50 excitation periods.

$\delta$: deviation from the measured values

## 5. Conclusion

This work presented a systematic investigation of the transient propagation characteristics of a magnetoelectric SAW sensor based on a multilayer Love-wave delay line using complementary time- and frequency-domain analyses. The propagation delay, phase velocity, and group velocity were directly extracted from the transient response within a 3D FEM framework, with the results cross-validated against independent frequency-domain dispersion analysis and experimental measurements. At the operating frequency of 146 MHz, a phase velocity of 4271 m/s and a group velocity of 3419 m/s were obtained. The difference between these two velocities demonstrates the dispersive nature of the Love wave and is consistent with the dispersion characteristics obtained independently from the frequency-domain analysis, where a phase velocity of 4105 m/s and a group velocity of 3405 m/s were calculated.

The simulated delay time at 146 MHz was 1.321 µs, compared with 1.357 µs measured experimentally, corresponding to a deviation of 2.7%. The group velocity obtained from wavefront tracking also showed good agreement with the experimental value, with a deviation of approximately 0.9%. These results demonstrate the consistency of the time-domain approach with the frequency-domain analysis and experimental measurements, confirming the accuracy of the numerical model in reproducing the experimentally observed transient response.

The present study focuses exclusively on the SAW propagation characteristics and transient response of the sensor, rather than its magnetic-field sensing performance. The FeCoSiB magnetostrictive layer is included as a structural constituent of the fabricated device in the current study. The resulting magnetic-field-dependent response including the magnetic hysteresis of the film will be investigated in subsequent studies.

## Appendix A. Fabrication Method

The SAW sensor is based on a piezoelectric ST-cut quartz substrate. IDTs were fabricated by DC magnetron sputtering using a VON ARDENNE CS 730S cluster system. They consist of a 350-nm-thick Au film sandwiched between 10-nm-thick chromium adhesion layers. Patterning of the electrodes was carried out using a lift-off process. Arranged in a delay-line configuration, the IDTs comprise 25 double-finger electrode pairs with a periodicity of $p$=28 μm. A $SiO_2$ layer with a thickness of 4.55 μm was deposited by plasma-enhanced chemical vapor deposition (PECVD) to confine the acoustic energy near the surface. The guiding layer was subsequently patterned using inductively coupled plasma reactive ion etching (ICP-RIE). Together with the perpendicular orientation of the delay line relative to the crystallographic $x$-axis of the quartz substrate, this configuration results in a dominant shear-horizontal surface mode, known as a Love wave. The sensing region consists of a 200-nm-thick amorphous magnetostrictive thin film of $(Fe_{90}Co_{10})_{78}Si_{12}B_{10}$ deposited by RF magnetron sputtering. The rectangular magnetic film pattern, measuring 3.8 mm along the delay line and 3 mm perpendicular to it, is defined between the IDTs using a lift-off process. Beneath the magnetostrictive layer, a 10-nm-thick tantalum film was deposited to improve adhesion, while an additional tantalum layer was applied on top to provide protection against corrosion. During deposition, an external magnetic field was applied to induce a uniaxial magnetic anisotropy aligned with the propagation direction of the Love wave.

## Appendix B. Material Parameters

In the model, ST-cut quartz is used as the piezoelectric substrate. Its material parameters are derived from those of α-quartz [41] by applying a 42.75° rotation about the $x$-axis, followed by a 90° rotation about the $z$-axis. The resulting parameters are given as follows:

$$\mathbf{C}_{\mathrm{ST-quartz}} = \begin{pmatrix} 96.6 & 27.5 & -4.8 & 0 & 13.4 & 0 \\ 27.5 & 86.7 & -8.6 & 0 & 1.1 & 0 \\ -4.8 & -8.6 & 130.7 & 0 & -1.8 & 0 \\ 0 & 0 & 0 & 30.4 & 0 & 7.6 \\ 13.4 & 1.1 & -1.8 & 0 & 41.2 & 0 \\ 0 & 0 & 0 & 7.6 & 0 & 67.5 \end{pmatrix} \mathrm{GPa},$$

$$\mathbf{e}_{\mathrm{ST-quartz}} = \begin{pmatrix} 0 & 0 & 0 & -0.07 & 0 & 0.10 \\ 0.04 & -0.17 & 0.13 & 0 & -0.08 & 0 \\ 0 & 0 & 0 & 0.07 & 0 & -0.11 \end{pmatrix} \mathrm{C/m^2},$$

$$\boldsymbol{\varepsilon}_{\mathbf{r}_{\mathrm{ST-quartz}}} = \begin{pmatrix} 4.5 & 0 & 0.1 \\ 0 & 4.4 & 0 \\ 0.1 & 0 & 4.5 \end{pmatrix},$$

$$\rho_{\mathrm{ST-quartz}} = 2651\ \mathrm{kg/m^3},$$

where $\mathbf{C}_{\text{ST-quartz}}$, $\mathbf{e}_{\text{ST-quartz}}$, $\boldsymbol{\varepsilon}_{\mathbf{r}\text{ST-quartz}}$ and $\rho_{\text{ST-quartz}}$ denote the stiffness tensor, piezoelectric coupling tensor, relative permittivity, and mass density of ST-cut quartz, respectively.

The guiding layer is composed of $SiO_2$ with the following parameters [42]:

$$\mathbf{C}_{\mathrm{SiO_2}} = \begin{pmatrix} 67.9 & 17.5 & 17.5 & 0 & 0 & 0 \\ 17.5 & 67.9 & 17.5 & 0 & 0 & 0 \\ 17.5 & 17.5 & 67.9 & 0 & 0 & 0 \\ 0 & 0 & 0 & 25.2 & 0 & 0 \\ 0 & 0 & 0 & 0 & 25.2 & 0 \\ 0 & 0 & 0 & 0 & 0 & 25.2 \end{pmatrix} \mathrm{GPa},$$

$$\mathbf{e}_{\mathrm{SiO_2}} = \begin{pmatrix} 0 & 0 & 0 & 0 & 0 & 0 \\ 0 & 0 & 0 & 0 & 0 & 0 \\ 0 & 0 & 0 & 0 & 0 & 0 \end{pmatrix} \mathrm{C/m^2},$$

$$\boldsymbol{\varepsilon}_{\mathbf{r}_{\mathrm{SiO_2}}} = \begin{pmatrix} 2.2 & 0 & 0 \\ 0 & 2.2 & 0 \\ 0 & 0 & 2.2 \end{pmatrix},$$

$$\rho_{\mathrm{SiO_2}} = 2144\ \mathrm{kg/m^3}.$$

The magnetostrictive layer consists of FeCoSiB, characterized by a mass density of $\rho$=7250 kg/m$^3$, a Young's modulus of $E_0$=150 GPa, and a Poisson's ratio of $\upsilon$=0.3 [33]. According to the magnetoelastic model described elsewhere [25], the elastic parameters of the FeCoSiB layer are identical under zero-field and saturation conditions.

The adhesion and protection layers used in the fabricated device were not included in the numerical model to avoid unnecessary model complexity, as their thickness is negligible compared with the thicknesses of the functional layers and their influence on the acoustic response is expected to be minor.

**Appendix C. Measurement Setup**

The measurement of the delay time is conducted through the utilization of a configuration comprising two synchronously operated PC-based measurement cards. The SAW is driven by generating a signal with a specified phase, amplitude, and frequency, which is then output via two channels of an M4i.6622-x8 16-bit high-speed arbitrary waveform generator from *Spectrum Instrumentation*. One signal is routed directly to the data acquisition system (reference), whereas the other passes through the SAW to the second acquisition channel (measurement). Data acquisition is facilitated by an M4i.4451-x8 14-bit high-speed digitizer, also manufactured by *Spectrum Instrumentation*. The M4i.4451-x8 is utilized as the clock source. It is noteworthy that both cards operate on a shared clock, facilitating synchronized signal generation and recording at a rate of 500 MS/s. The generated signal has a power of 0 dBm. The recording process is initiated automatically upon the detection of a signal level of 50 mV on the reference channel, and the recording covers 256 pre-trigger data points. The sensor under investigation is placed in a magnetically shielded environment throughout the entire measurement.

The recorded measurement signal is first up-sampled by a factor of 10 using cubic spline interpolation. This is possible because the frequencies under investigation are lower than the Nyquist frequency (250 MHz) of the data acquisition system. This up-sampling process improves the temporal resolution to 200 ps. The initial signal generation recorded on the reference channel defines $t$=0 s. Furthermore, the recorded measurement signal is corrected for the influence of the cables used (2×3 m CLF100) in terms of both its delay and amplitude. All measurements were performed in a magnetic shield with no applied bias field. This approach was justified by the fact that the SAW velocity under zero-field conditions is equivalent to that under magnetic saturation. This simplifies the measurement setup and reduces external noise sources.

## Acknowledgements

This work was supported by the German Research Foundation (Deutsche Forschungsgemeinschaft, DFG) through the Collaborative Research Centre CRC 1261 Magnetoelectric Sensors: From Composite Materials to Biomagnetic Diagnostics. The authors would like to express their gratitude to Project DEAL for funding open access publication.

## References

[1] R. Weiss *et al.*, “Magnetic sensors-A review and recent technologies,” *Eng. Res. Express*, vol. 3, no. 2, p. 022005, Jun. 2021, DOI: 10.1088/2631-8695/AC0838.

[2] X. Bai, K. Wen, D. Peng, S. Liu, and L. Luo, “Atomic magnetometers and their application in industry,” *Front. Phys.*, vol. 11, p. 1212368, Jun. 2023, DOI: 10.3389/FPHY.2023.1212368.

[3] K. Zhao *et al.*, “Magnetic Field-Based Non-Destructive Testing Techniques for Battery Diagnostics,” *Adv. Energy Mater.*, vol. 15, no. 10, p. 2404295, Mar. 2025, DOI: 10.1002/AENM.202404295.

[4] P. Lipovský, P. Ripka, K. Draganová, and A. Polakovič, “Precise Measurement of Magnetic Field for Geophysical, Space, and Security Applications,” *IEEE Sens. J.*, vol. 25, no. 13, pp. 23998–24007, 2025, DOI: 10.1109/JSEN.2025.3570654.

[5] T. Sander, A. Jodko-Władzińska, S. Hartwig, R. Brühl, and T. Middelmann, “Optically pumped magnetometers enable a new level of biomagnetic measurements,” *Adv. Opt. Technol.*, vol. 9, no. 5, pp. 247–251, Oct. 2020, DOI: 10.1515/AOT-2020-0027.

[6] K. Zhu and A. Kiourti, “A Review of Magnetic Field Emissions from the Human Body: Sources, Sensors, and Uses,” *IEEE Open J. Antennas Propag.*, vol. 3, pp. 732–744, 2022, DOI: 10.1109/OJAP.2022.3186643.

[7] B. Bao, Y. Hua, R. Wang, and D. Li, “Quantum-Based Magnetic Field Sensors for Biosensing,” *Adv. Quantum Technol.*, vol. 6, no. 5, p. 2200146, May 2023, DOI: 10.1002/QUTE.202200146.

[8] J. Hoffmann *et al.*, “Pilot Study: Step Width Estimation with Body-Worn Magnetoelectric Sensors,” *Sensors*, vol. 25, no. 11, p. 3390, May 2025, DOI: 10.3390/S25113390.
[9] G. Barbieri *et al.*, “Unshielded Portable Gradiometer Setup for Magnetic Nanoparticles Characterization in Flow Condition,” *IEEE Sens. J.*, vol. 26, no. 3, pp. 3812–3822, Feb. 2026, DOI: 10.1109/JSEN.2025.3641840.
[10] V. Polewczyk *et al.*, “Unipolar and bipolar high-magnetic-field sensors based on surface acoustic wave resonators,” *Phys. Rev. Appl.*, vol. 8, no. 2, p. 024001, Aug. 2017, DOI: 10.1103/PHYSREVAPPLIED.8.024001.
[11] A. Kittmann *et al.*, “Wide Band Low Noise Love Wave Magnetic Field Sensor System,” *Sci. Rep.*, vol. 8, no. 1, pp. 1–10, Jan. 2018, DOI: 10.1038/s41598-017-18441-4.
[12] W. Wang, Y. Jia, X. Xue, Y. Liang, and Z. Du, “Grating-patterned FeCo coated surface acoustic wave device for sensing magnetic field,” *AIP Adv.*, vol. 8, no. 1, Jan. 2018, DOI: 10.1063/1.5012579/990884.
[13] X. Liu *et al.*, “Self-biased vector magnetic sensor based on a Love-type surface acoustic wave resonator,” *Appl. Phys. Lett.*, vol. 113, no. 8, Aug. 2018, DOI: 10.1063/1.5044478/37057.
[14] J. Schmalz *et al.*, “Multi-Mode Love-Wave SAW Magnetic-Field Sensors,” *Sensors*, vol. 20, no. 12, p. 3421, Jun. 2020, DOI: 10.3390/S20123421.
[15] J. M. Meyer *et al.*, “Thin-Film-Based SAW Magnetic Field Sensors,” *Sensors*, vol. 21, no. 24, p. 8166, Dec. 2021, DOI: 10.3390/S21248166.
[16] V. Schell *et al.*, “Exchange biased surface acoustic wave magnetic field sensors,” *Sci. Rep.*, vol. 13, no. 1, pp. 8446-, May 2023, DOI: 10.1038/s41598-023-35525-6.
[17] J. D. Livingston, “Magnetomechanical properties of amorphous metals,” *Phys. Status Solidi A*, vol. 70, no. 2, pp. 591–596, Apr. 1982, DOI: 10.1002/PSSA.2210700228.
[18] C. Dong *et al.*, “Characterization of magnetomechanical properties in FeGaB thin films,” *Appl. Phys. Lett.*, vol. 113, no. 26, Dec. 2018, DOI: 10.1063/1.5065486/36361.
[19] X. Liang *et al.*, “A Review of Thin-Film Magnetoelastic Materials for Magnetoelectric Applications,” *Sensors*, vol. 20, no. 5, p. 1532, Mar. 2020, DOI: 10.3390/S20051532.
[20] B. Spetzler *et al.*, “Magnetoelastic Coupling and Delta-E Effect in Magnetoelectric Torsion Mode Resonators,” *Sensors*, vol. 21, no. 6, p. 2022, Mar. 2021, DOI: 10.3390/S21062022.
[21] D. C. Webb, D. W. Forester, A. K. Ganguly, and C. Vittoria, “Applications of amorphous magnetic-layers in surface-acoustic-wave devices,” *IEEE Trans. Magn.*, vol. 15, no. 6, pp. 1410–1415, 1979, DOI: 10.1109/TMAG.1979.1060442.
[22] M. Yamaguchi, M. Naoe, and H. Kogo, “Variable saw delay line using amorphous TbFe2 film,” *IEEE Trans. Magn.*, vol. 16, no. 5, pp. 916–918, 1980, DOI: 10.1109/TMAG.1980.1060738.
[23] C. Müller *et al.*, “Imaging of Love Waves and Their Interaction with Magnetic Domain Walls in Magnetoelectric Magnetic Field Sensors,” *Adv. Electron. Mater.*, vol. 8, no. 6, p. 2200033, Jun. 2022, DOI: 10.1002/AELM.202200033.
[24] J. D. Aguilera *et al.*, “Anomalies in the magnetostrictive modulation of love surface acoustic waves,” *AIP Adv.*, vol. 14, no. 2, Feb. 2024, DOI: 10.1063/9.0000749/3261889.
[25] M. Samadi, J. Schmalz, J. M. Meyer, F. Lofink, and M. Gerken, “Phononic-Crystal-Based SAW Magnetic-Field Sensors,” *Micromachines*, vol. 14, no. 11, p. 2130, Nov. 2023, DOI: 10.3390/MI14112130.
[26] M. Samadi *et al.*, “Modeling of High-Sensitivity SAW Magnetic Field Sensors with Au-SiO2 Phononic Crystals,” *Adv. Sensor Res.*, vol. 4, no. 5, p. 2500008, May 2025, DOI: 10.1002/ADSR.202500008.
[27] C. Chen *et al.*, “Highly Sensitive Surface Acoustic Wave Magnetic Field Sensor Based on a Strongly Coupled Magnon-Phonon System,” *Adv. Funct. Mater.*, vol. 36, no. 19, p. e18999, Mar. 2026, DOI: 10.1002/ADFM.202518999.
[28] K. Hashimoto, “Surface Acoustic Wave Devices in Telecommunications,” vol. 116, Berlin: Springer-Verlag, 2000, DOI: 10.1007/978-3-662-04223-6.
[29] M. Hofer *et al.*, “Finite-element simulation of wave propagation in periodic piezoelectric SAW structures,” *IEEE Trans. Ultrason. Ferroelectr. Freq. Control*, vol. 53, no. 6, pp. 1192–1200, Jun. 2006, DOI: 10.1109/TUFFC.2006.1642518.
[30] S. Trivedi and H. B. Nemade, “Finite element simulation of a highly sensitive SH-SAW delay line sensor with $SiO_2$ micro-ridges,” *Microsyst. Technol.*, vol. 24, no. 8, pp. 3537–3547, Feb. 2018, DOI: 10.1007/S00542-018-3772-Y.
[31] A. S. Koigerov and A. V. Korlyakov, “Finite Element Modeling of Surface Acoustic Wave Devices Using COMSOL,” *Russ. Microelectron.*, vol. 51, no. 4, pp. 226–235, Jul. 2022, DOI: 10.1134/S1063739722040072.
[32] J. Labrenz *et al.*, “Frequency Response of SAW Delay Line Magnetic Field/Current Sensor,” *IEEE Sens. Lett.*, vol. 3, no. 10, Oct. 2019, DOI: 10.1109/LSENS.2019.2943129.
[33] C. Cordier and C. Dolabdjian, “Modeling of Delta-E Effect Magnetic Field Sensors,” *IEEE Sens. J.*, vol. 23, no. 3, pp. 2014–2020, Feb. 2023, DOI: 10.1109/JSEN.2022.3227592.

[34] J. Schmalz, E. Spetzler, J. McCord, and M. Gerken, “Investigation of Unwanted Oscillations of Electrically Modulated Magnetoelectric Cantilever Sensors,” *Sensors*, vol. 23, no. 11, p. 5012, May 2023, DOI: 10.3390/S23115012.

[35] J. Yang, “An Introduction to the Theory of Piezoelectricity,” vol. 9, Springer, 2018, DOI: 10.1007/978-3-030-03137-4.

[36] IEEE Standard on Piezoelectricity, Standard 176-1987, 1987.

[37] COMSOL Multiphysics® v. 6.4, COMSOL AB, Stockholm, Sweden.

[38] H. Wolframm *et al.*, “Implementation of a Segmented PLL for Readout of Highly Sensitive SAW Magnetometers,” *Curr. Dir. Biomed. Eng.*, vol. 11, no. 1, pp. 214–217, Sep. 2025, DOI: 10.1515/CDBME-2025-0155.

[39] G. Kovacs, G. W. Lubking, M. J. Vellekoop, and A. Venema, “Love waves for (bio)chemical sensing in liquids,” *Proc. IEEE Ultrason. Symp.*, pp. 281–285, Oct. 1992, DOI: 10.1109/ULTSYM.1992.275995.

[40] F. Martin, G. McHale, and M. I. Newton, “Experimental Study of Love Wave Sensor Response by Phase and Group Velocity Measurement,” *IEEE Sens. J.*, vol. 4, no. 2, pp. 216–220, Apr. 2004, DOI: 10.1109/JSEN.2004.823674.

[41] R. Bechmann, “Elastic and Piezoelectric Constants of Alpha-Quartz,” *Phys. Rev.*, vol. 110, no. 5, p. 1060, Jun. 1958, DOI: 10.1103/PhysRev.110.1060.

[42] M. Knapp *et al.*, “Accurate characterization of SiO2 thin films using surface acoustic waves,” *IEEE Trans. Ultrason. Ferroelectr. Freq. Control*, vol. 62, no. 4, pp. 736–743, Apr. 2015, DOI: 10.1109/TUFFC.2014.006921.